\documentclass[showpacs,preprintnumbers,amsmath,amssymb,floatfix,12pt]{revtex4}

\usepackage{graphicx}
\usepackage{dcolumn}
\usepackage{bm}

\begin{document}

%

\let\a=\alpha      \let\b=\beta       \let\c=\chi        \let\d=\delta
\let\e=\varepsilon \let\f=\varphi     \let\g=\gamma      \let\h=\eta
\let\k=\kappa      \let\l=\lambda     \let\m=\mu
\let\o=\omega      \let\r=\varrho     \let\s=\sigma
\let\t=\tau        \let\th=\vartheta  \let\y=\upsilon    \let\x=\xi
\let\z=\zeta       \let\io=\iota      \let\vp=\varpi     \let\ro=\rho
\let\ph=\phi       \let\ep=\epsilon   \let\te=\theta
\let\n=\nu
\let\D=\Delta   \let\F=\Phi    \let\G=\Gamma  \let\L=\Lambda
\let\O=\Omega   \let\P=\Pi     \let\Ps=\Psi   \let\Si=\Sigma
\let\Th=\Theta  \let\X=\Xi     \let\Y=\Upsilon

%

%

\def\cA{{\cal A}}                \def\cB{{\cal B}}
\def\cC{{\cal C}}                \def\cD{{\cal D}}
\def\cE{{\cal E}}                \def\cF{{\cal F}}
\def\cG{{\cal G}}                \def\cH{{\cal H}}
\def\cI{{\cal I}}                \def\cJ{{\cal J}}
\def\cK{{\cal K}}                \def\cL{{\cal L}}
\def\cM{{\cal M}}                \def\cN{{\cal N}}
\def\cO{{\cal O}}                \def\cP{{\cal P}}
\def\cQ{{\cal Q}}                \def\cR{{\cal R}}
\def\cS{{\cal S}}                \def\cT{{\cal T}}
\def\cU{{\cal U}}                \def\cV{{\cal V}}
\def\cW{{\cal W}}                \def\cX{{\cal X}}
\def\cY{{\cal Y}}                \def\cZ{{\cal Z}}

%

\newcommand{\Ns}{N\hspace{-4.7mm}\not\hspace{2.7mm}}
\newcommand{\qs}{q\hspace{-3.7mm}\not\hspace{3.4mm}}
\newcommand{\ps}{p\hspace{-3.3mm}\not\hspace{1.2mm}}
\newcommand{\ks}{k\hspace{-3.3mm}\not\hspace{1.2mm}}
\newcommand{\des}{\partial\hspace{-4.mm}\not\hspace{2.5mm}}
\newcommand{\desco}{D\hspace{-4mm}\not\hspace{2mm}}



\title{\boldmath Anomalous gauge boson couplings, 125 GeV Higgs and singlet scalar dark matter }

\author{Namit Mahajan
}
\email{nmahajan@prl.res.in}
\affiliation{
 Theoretical Physics Division, Physical Research Laboratory, Navrangpura, Ahmedabad
380 009, India
}


\begin{abstract}
The recently observed Higgs like resonance at 125 GeV shows an enhanced rate in the
diphoton channel, while being roughly consistent with the standard model expectation for the $WW^*$ and $ZZ^*$ channels. Such an enhancement is possible due to anomalous gauge boson couplings.
We explore this feature within a minimal extension of the standard model, where a singlet scalar
is introduced which also plays the role of the dark matter candidate. It is argued that such
a minimal scenario, without new charged particles, can in principle lead to the desired enhancement
of the diphoton rate via the induced anomalous gauge couplings, and at the same time improve the stability of the electroweak vacuum.
\end{abstract}

\pacs{
}
\maketitle


The ATLAS \cite{:2012gk} and CMS \cite{:2012gu} experiments have both observed a Higgs \cite{Higgs:1964ia} like resonance, with a mass $~125$ GeV. Observation of the two photon
final state rules out spin-1 as a plausible option for such a state \cite{Yang:1950rg} while strictly speaking spin-2 still remains a viable option. In the Standard Model (SM), the Higgs boson couples to all the particles with a strength proportional to their masses. Therefore, there is no coupling to photons at the tree level but is generated at the one loop level \cite{Ellis:1975ap}. Within SM, the $H\to\gamma\gamma$ rate is dominated by the W-boson loops (see \cite{Djouadi:2005gi} for a detailed compilation of results including higher order QCD and eletroweak corrections). The rate in the diphoton channel appears to be $1.5$-$2$ times the SM expectation, implying $~2\sigma$ deviation. If this result survives when more data is added, this would undoubtly be a signature of physics beyond SM. The rates for $H\to ZZ^*,\, WW^*$ are consistent with the SM values while $H\to bb,\,\tau\tau$ seem to be low but at present these channels are not expected to have a high sensitivity. 
There has been a flury of activity exploring various possibilities and implications of this result \cite{Low:2012rj}. Many possibilities exist to enhance the diphoton rate, typically requiring new charged particles going around in the loop. It has also been suggested that once the theoretical uncertainties are taken into account properly, this $2\sigma$ deviation becomes a $1\sigma$ deviation \cite{Baglio:2012et}. In any case, within the present errors (including theoretical uncertainties), it may not be wrong to say that a modest $20$-$30\%$ enhancement is still allowed.

Over the years, particularly via the LEP experiments, the gauge boson couplings to fermions has been tested with great precision. The bosonic sector of the SM gauge theory however has not been tested to the same precision. The presence of new physics beyond the standard model at a generic scale $\Lambda$ leads to higher dimensional operators when the massive degrees of freedom are integrated out leading to an effective Lagrangian
\begin{equation}
{\mathcal{L}}_{\textrm eff} = {\mathcal{L}}_{0} + \sum_{n=5}\frac{\alpha_i}{\Lambda^{n-4}}O^{(n)}_i
\end{equation}
where $O^{(n)}_i$ are non-renormalizable operators. One expects that a dimension six operator generated at one loop level is of the order $(v/\Lambda)^2(\alpha_i/16\pi^2)$, where $v$ is the SM Higgs vacuum expectation value. Within SM, the three and four point vertices between the gauge boson are completely determined following the local $SU(2)_L\times U(1)_Y$ gauge symmetry. Further, the gauge structure of SM relates the three and four point gauge boson couplings: $g_{WWVV} = g^2_{WWV}$.
The presence of new physics could drastically change such a conclusion.
The $WW\gamma$ vertex, see for example \cite{Gounaris:1996rz}, is parameterized by the effective Lagrangian (CP conserving terms only)
\begin{equation}
\Delta{\mathcal{L}}_{\textrm WW\gamma} = -ie\left(g_1^{\gamma}(W_{\mu\nu}^+W^{-\mu}-W_{\mu\nu}^-W^{+\mu})A^{\nu} + \kappa_{\gamma}W^+_{\mu}W^-_{\nu}F^{\mu\nu} + \frac{\lambda_{\gamma}}{m_W^2}W_{\mu}^{+\nu}W_{\nu}^{-\rho}F_{\rho}^{\mu}\right)
\end{equation}
Within SM, the tree level couplings are given by $g_1^{\gamma} = \kappa_{\gamma} = 1$ while $\lambda_{\gamma} = 0$. $g_1^{\gamma}$ corresponds to the electric charge of the W-boson while $\kappa_{\gamma}$ and $\lambda_{\gamma}$ can be identified with the anomalous magnetic and quadrupole moments respectively:
\begin{equation}
\mu_W = \frac{e}{2m_W}(1+\Delta\kappa_{\gamma} + \lambda_{\gamma}) \hskip 1.5cm 
Q_W = \frac{e}{m_W^2}(\Delta\kappa_{\gamma} - \lambda_{\gamma})
\end{equation}
where $\Delta\kappa_{\gamma} = \kappa_{\gamma} - 1$ parameterizes the deviation from the SM tree level value.
It turns out that LEP data \cite{Alcaraz:2006mx} provides the best limits on the anomalous couplings:
\begin{equation}
-0.098 < \Delta\kappa_{\gamma} < 0.101 \hskip 1.5cm -0.044 < \lambda_{\gamma} < 0.047
\end{equation}
In a similar fashion, anomalous contributions to quartic gauge boson couplings can be parameterized \cite{Belanger:1992qh} and similar constraints obtained.

We now turn our attention to a very simple extension of SM, namely adding a single scalar to the SM field content \cite{Silveira:1985rk}. For simplicity we consider a real scalar. Generalization to complex scalar is straightforward.
Such a simple model has many attractive features. It is known that within SM, for Higgs mass around 125 GeV, the Higgs quartic coupling becomes negative at some large energy much before the Planck scale \cite{Isidori:2001bm}. This vacuum instability crucially depends on the top quark mass and the value of the strong coupling constant \cite{Alekhin:2012py}. Inclusion of additional scalar particle is generically expected to provide a positive contribution to the running of the Higgs quartic coupling, compensating the large negative contribution due to the top quark loop, thereby improving the stability of the vacuum \cite{Gonderinger:2009jp}, \cite{Gonderinger:2012rd}. SM augmented by a singlet scalar is an elegant model for cold dark matter \cite{McDonald:1993ex}. The Lagrangian for the model reads
\begin{equation}
{\mathcal{L}} = {\mathcal{L}}_{\textrm SM} + \frac{1}{2}(\partial_{\mu}S)^2 - \frac{1}{2}m_0^2S^2 - \frac{1}{4!}\eta S^4 - \frac{1}{2}\rho S^2\Phi^{\dag}\Phi
\end{equation}
where $\Phi$ is the SM Higgs doublet and $S$ is the singlet real scalar. Imposing a ${\mathbb{Z}}_2$ symmetry such that $S\to -S$ with all the SM fields unchanged ensures that there are no odd terms in the field $S$. This ensures the stability of the singlet field and makes it a viable dark matter candidate. Further, to avoid cosmological problems, it is necessary that $S$ does not acquire vacuum expectation value. As is evident from the above Lagrangian, $\langle S\rangle = 0$ implies that there is no mixing between $S$ and the physical Higgs field after electroweak symmetry breaking. Therefore $m_0^2 > 0$. After electroweak symmetry breaking, the relevant part of the Lagrangian in the unitary gauge is
\begin{equation}
{\mathcal{L}} = -\frac{1}{2}\underbrace{(m_0^2 + \frac{\rho v^2}{2})}_{m_S^2} S^2 - \frac{1}{2}\rho vHS^2 - \frac{1}{4}\rho H^2S^2 - \frac{1}{4}\eta S^4 \label{lag}
\end{equation} 
The Higgs mass is related to the Higgs quartic coupling $\lambda$ via the usual relation: $m_H = \sqrt{2\lambda v}$. In the following we assume that $m_S > m_H$.
Stability of the potential demands that
\begin{equation}
\lambda > 0 \hskip 1cm \eta > 0 \hskip 1cm \eta\lambda > \rho^2
\end{equation}
while tree level perturbative unitarity requires
\begin{equation}
m_H^2 < \frac{8\pi}{3}v^2 \hskip 1cm \eta < 8\pi \hskip 1cm \rho < 8\pi
\end{equation} 

Self interactions of the singlet field, having a mean free path $\lambda_S$, lead to an almost model independent result \cite{Bento:2000ah}:
\begin{equation}
\sigma_{SS} \equiv \sigma(SS\to SS) = 8.1\times 10^{-25}\left(\frac{\lambda_S}{Mpc}\right)^{-1}\,{\mathrm cm}^2/{\mathrm GeV}
\end{equation}
where for non-relativistic dark matter particles ($s\sim (2m_S)^2$)
\begin{equation}
\sigma_{SS} = \frac{\eta^2}{16\pi s} \sim \frac{\eta^2}{64\pi m_S^2}
\end{equation}
The singlet self coupling $\eta$ is not too well constrained and could in principle be somewhat large, ${\mathcal{O}}(1)$ and still consistent with the observed relic density. 

For $m_S > m_H$, $SS\to HH$ is kinematically allowed and we have
\begin{equation}
\sigma_{\textrm ann}v_{\mathrm rel} \sim \frac{\rho^2}{4\pi m_S^2}
\end{equation}
and
\begin{equation}
\Omega_Sh^2 \sim \frac{x_f 10^9}{g_*^{1/2}M_{Pl}\langle \sigma_{\mathrm ann}v_{\mathrm rel}\rangle}
\end{equation}
Equating this with the observed relic density \cite{Komatsu:2010fb}, yields $\rho \sim$ 0.02-0.05.
The second term in Eq.(\ref{lag}) will be responsible for scattering of S off nucleons in the direct detection experiments. The spin-independent elastic scattering off a nucleon is given by
\begin{equation}
\sigma_{\mathrm el}(\mathrm nucleon) \sim \rho^2 (20\times 10^{-42})\left(\frac{50 {\mathrm GeV}}{m_S}\right)^2\left(\frac{100 {\mathrm GeV}}{m_H}\right)^2 \,{\mathrm cm}^2
\end{equation}
which for $m_H =125$ GeV and $m_S \sim 150$ GeV implies $\sigma_{\mathrm el}(\mathrm nucleon) \sim 10^{-45}$ which is consistent with the recent Xenon exclusion limits \cite{Aprile:2012nq}.

We therefore find that the minimal extension of SM by adding just one singlet scalar seems to be totally consistent with all the available constraints and provides a viable and economical dark matter candidate. The vacuum stability and perturbativity of such a model has been studied in detail \cite{Gonderinger:2009jp}. The model seems to be stable upto the Planck scale for the choice of parameters obtained above namely $m_H=125$ GeV, $m_S \sim 150$ GeV, $\eta \sim 1$ and $\rho \sim 0.02$, and simultaneously being consistent with WMAP results on the dark matter relic density. As mentioned before, the additional scalar degree of freedom contributes to the running of the quartic Higgs coupling such that it partly compenstaes for the large negative top quark contribution, thereby ensuring that the Higgs potential doesnot turn negative till values close to the Planck scale. The smallness of $\rho$ also ensures that one loop corrections to the Higgs effective potential are still dominated by the SM degrees of freedom and the singlet only corrects it marginally.

We now turn to the issue of anomalous gauge boson couplings in the present context. To motivate and as a proof of existence of such a possibility let us recall the arguments of \cite{Hill:1987ea}. These authors considered a simple model where a real singlet is added to the SM but for simplicity the interaction considered had the form $S\Phi^{\dag}\Phi$ and both the SM Higgs and the singlet aqcuire vacuum expctation values. The authors carefully study the model in detail to one loop and find that the new contributions to $\rho$ parameter remain small while generating sizeable anomalous gauge boson couplings. Various limits are discussed and it has been stressed that if the two mass eigenvalues are similar in size, then the effects essentially depend on the Higgs mass. However, in the other limit when the ratio of the singlet to Higgs mass grows with the self coupling, new corrections to three and four point gauge boson vertices arise. The present model is different from the one considered in \cite{Hill:1987ea} in one essential way. The singlet field in the present context does not acquire a vacuum expectation value and therefor there is no mixing with the Higgs field at the tree level. This however is not a severe drawback. One could consider explicit soft breaking or gauging the extra ${\mathbb{Z}}_2$ symmetry, thereby letting $S$ acquire (even a large) vecuum expectation value (denoted by $v_S$) and still avoid cosmological problems. The other qualitative as well as quanitative features relatd to dark matter phenomenology are not expected to be significantly altered due to this soft breaking.
Consider the case of large $v_S (>v)$
\begin{equation}
\tan 2\theta \simeq -\frac{\rho v}{\eta v_S} \hskip 1cm m_{\mathrm light}^2 \simeq 2\left(\lambda - \frac{\rho^2}{4\eta} \right)v^2
\hskip 1cm m_{\mathrm heavy}^2 \simeq 2\eta v_S^2 + \frac{\rho^2 v^2}{4\eta}
\end{equation}
For $v_S > v$, and $\eta$ somewhat large ($> \rho$), the mixing angle between the two states becomes small and the light state effectively becomes the SM Higgs while the heavier state tends to be a pure singlet. As $v_S\to\infty$, both $m_S^2$ and $m_H^2$ grow while their ratio is held fixed to the ratio of $\eta$ and $\rho$. This is very similar to the requirements in \cite{Hill:1987ea} needed to induce sizeable anomalous gauge boson couplings. 
 The detailed evaluation of the anomalous couplings is beyond the scope of the present study and will be presented elsewhere. Rough estimates indicate that anomalous couplings $\sim {\mathrm few}\times 10^{-3}$ are easily possible. Also, for the sake of illustration, we have considered $m_S \sim 150$ GeV but the singlet scalar could be somewhat heavier, particularly when considering the scenario where ${\mathbb{Z}}_2$ symmetry is broken. This would imply additional logarithmic corrections that now depend on the ratio of the two masses. Therefore the values of anomalous couplings like the one mentioned above do not seem to be unrealistic, although these are only rough estimates which should be corroborated by a complete calculation. For the present, we assume that the above estimate holds true.

It has been explicitly shown \cite{De Rujula:1991se} that a loop induced process like $H\to\gamma\gamma$ is very sensitive to such anomalous couplings and an enhancement as large as an order of magnitude over the SM expectation is possible. Explicit expressions for the additional contributions to the diphoton amplitude are somewhat lengthy and are not reprodcued here.
Within SM, the typical size of the anomalous couplings can be estimated to be $< 10^{-3}$. In particular, the CP violating couplings are not expected to appear upto two loop level, similar to the electric dipole moments. Physics beyond SM could lead to large anomalous couplings and therefore can have perceptible effects. The magnitude of the anomalous couplings could be estimated given a specific model beyond SM. This could in turn be used to constrain the parameters of the model.
As discussed before, we are here looking for a modest enhancement (say $\sim 20\%$) of the Higgs to diphoton rate, which translates to $10\%$ enhancement at the amplitude level, which is easily achieved.
Anomalous gauge couplings would also contribute to other processes like $H\to Z\gamma$. However, compared to $H\to\gamma\gamma$, the anomalous contributions are only marginal. Muon anomalous magnetic moment provides one more laboratory to test the sensitivity of the anomalous couplings. Following \cite{Boudjema:1990dv}, we find that the absolute magnitude of the correction to muon anomalous moment due to anomalous gauge couplings of the size considered here can be non-negligible and will depend on the signs of the anomalous couplings. One can further verify that the contribution to the muon magnetic moment due to the scalar singlet is abysmally small \cite{Cheung:2012xb}. We have checked that assuming the above quoted rough values for the anomalous couplings, there are no dangerous contributions to  $b\to s\gamma$.

In this short note we have suggested that the scalar singlet dark matter model could indirectly lead to a sizeable enhancement of $H\to\gamma\gamma$ rate consistent with the recent observations by ATLAS and CMS collaborations. The model considered here is the simplest possible one and rough estimates for the induced anomalous gauge boson couplings have been obtained. These suffice to explain the observed enhancement. The model therefore turns out to be highly predictive and economical at the same time. It provides a viable dark matter candidate consistent with all available constraints, helps in improving the vacuum stability of the standard model upto very high energies. In the passing we also note that extending the scalar sector by adding a singlet could help in rescuing \cite{Lebedev:2012zw} the Higgs inflation scenario \cite{Bezrukov:2008ej} where in the early era, SM Higgs serves the role of the inflaton responsible for density fluctuations in the universe (see also \cite{Clark:2009dc} for a singlet scalar model of Higgs inflation and dark matter). It should be remarked that one may be forced to choose a variant of the simplest model considered here if one wishes to concretely determine the size and signs of anomalous couplings. The main message is that in the absence of any observation of physics beyond the standard model apart from tentalizing hints like enhanced $H\to\gamma\gamma$ rate, it may be premature to conclude that there is no new physics at the TeV scale besides the requirement of a dark matter candidate. In fact, the same dark matter candidate, even though neutral, could via indirect means end up enhancing the diphoton rate and also in principle, depending on the relative signs of various anomalous couplings, reduce the tension between the the observed and predicted values of muon anomalous magnetic moment. A detailed and systematic study of singlet scalar dark matter incorporating all these constraints will be presented elsewhere. We conclude that non-observation of new particles at the TeV scale, particularly electrically charged, and persistence of a largish diphoton rate could be strong indications towards a scenario similar to the one discussed here.



%

\end{document}